\documentclass[twocolumn]{jpsj2} 
\title{Redistribution of Electronic Charges in Spin-Peierls State in (TMTTF)$_2$AsF$_6$ Observed by $^{13}$C NMR}

\author{Shigeki \textsc{Fujiyama}\thanks{Present address: Department of Applied Physics, University of Tokyo, Hongo, Bunkyo-ku, Tokyo 113-8656, Japan, and CREST, Japan Science and Technology Corporation, Kawaguchi 332-0012, Japan, E-mail address: fujiyama@ap.t.u-tokyo.ac.jp}, Toshikazu \textsc{Nakamura}}

\inst{Institute for Molecular Science, Myodaiji, Okazaki 444-8585, Japan}

\abst{We report  $^{13}$C NMR spectra and nuclear spin lattice relaxation rate $1/T_1$ for a quasi-one-dimensional quarter-filled organic material (TMTTF)$_{2}$AsF$_{6}$, which undergoes charge ordering ($T_{\textrm{CO}}$ = 102 K) and spin-Peierls phase transitions ($T_{\textrm{SP}}$ = 14 K). The ratio of two $1/T_1$ for the charge accepting and donating TMTTF sites which grows from $T_\textrm{CO}$ finally saturates in approaching $T_{\textrm{SP}}$, indicating one spin correlation function even in the charge ordered state. Below $T_{\textrm{SP}}$, however, the doubly split NMR lines from inequivalently charged molecules merge into one line, originated from the variation in \textit{charge} densities. This shows that a rearrangement of the charge configuration occurs at $T_{\textrm{SP}}$.}

\kword{charge ordering, spin-Peierls transition, organic conductor, NMR}

\begin{document}
\maketitle

\section{Introduction}
\label{sec:Intro}
There has been considerable interest in correlated electrons in low dimensional systems. There,  interplay between spin, charge, and lattice degrees of freedom becomes relevant to the macroscopic physical properties, resulting in various electronic states. Recent developments on inorganic materials such as high-$T_\textrm{c}$ cuprates or manganites have revealed that electronic states can be highly inhomogeneous due to the complex interplay among these degrees of freedom.~\cite{Tranquada1995,Mori1998,Hanaguri2004}

Organic charge transfer salts (CTS) also provide model systems of strongly correlated low dimensional electrons.
Recent experimental and theoretical developments on the organic conductors have revealed that  in some compounds the electronic charges on individual cation molecules are varied (ordered), namely, inhomogeneous electronic states are realized.~\cite{Seo2004} Observations of anomalous dielectric relaxation, superstructures by X-ray diffraction (XRD), and shifts of Raman frequencies originating from the vibration of variedly charged molecules have indicated such inhomogeneous electronic states.~\cite{Nad1998,Pouget1997,Nogami2002,Yamamoto2002,Suzuki2004} Although recent theoretical developments have revealed that Coulomb repulsion among the electronic charges is indispensable to the charge ordering, various orderings have been proposed such as $2k_F$-spin density wave (SDW), $2k_F$-charge density wave (CDW), $4k_F$-CDW, and also the varieties of their coexistences.~\cite{Seo1997,Kobayashi1997,Ung1994,Mazumdar1999,Mazumdar2000,Clay2003,Hirsch1983,Penc1994,Riera2000}

It is well recognized that NMR technique has played a key role in elucidating the electronic states of organic conductors since the hyperfine field is proportional to the spin density of each nucleus. By this reason, NMR on the $^{13}$C labeled CTS is much more sensitive to the change of the electronic properties than that on $^{1}$H since the spin densities on the central carbons in organic CTS is richer.
In fact, several NMR measurements have so far revealed the occurrence of charge ordering in organic conductors. A line splitting of the $^{13}$C NMR spectra in the semiconducting temperature range was first demonstrated for a powder sample of quasi-one-dimensional (Q1D) system, (DI-DCNQI)$_2$Ag that shows imbalanced charged molecules.~\cite{Hiraki1998} 
For quasi-two-dimensional organic conductors, an anomalous line broadening originating from one carbon site and the corresponding differentiation in $1/T_1$ were reported for a single crystal of $\theta$-(BEDT-TTF)$_2$RbZn(SCN)$_4$.~\cite{Miyagawa2000}
These two studies clearly shows the charge ordering from the microscopic point of view.

Charge ordering has also been indicated in members in the family of (TMT$C$F)$_{2}$\textit{X} ($C$ = Se, S), also known as the Bechgaard salts and their sulphur analogues. 
Their physical properties have been extensively studied so far because the materials realize various ground states by modifying chalcogens ($C$) and/or anions ($X$).~\cite{Ishiguro1998} In addition to the chemical pressure effect that is controlled by the substitution of $C$ or $X$, it is well established that physically applied pressure can easily change the electronic state of this series of materials. Their macroscopic electronic properties are summarized in a pressure ($P$) vs.\ temperature ($T$) phase diagram.~\cite{Jerome1991} In this phase diagram the ground state varies as spin-Peierls state, antiferromagnet, SDW, and superconductivity, from low-$P$ side to high-$P$ side. The electronic properties are extensively investigated theoretically as well, based on models for quarter-filled Q1D electronic systems.~\cite{Schulz1994,Kishine2002}

The charge ordering  in TMTTF CTS was first indicated for the members that have antiferromagnetic ground states. The dependence of the $^{1}$H-NMR spectra on the direction of external magnetic fields for (TMTTF)$_2$Br and (TMTTF)$_2$SCN suggests that nodes of spin density exist on the TMTTF molecules.~\cite{Nakamura1997,Nakamura1995} A theoretical proposal including intersite repulsive interaction suggests an (up-0-down-0) magnetic structure as their ground state~\cite{Seo1997} and a recent $^{13}$C NMR study by the present authors on (TMTTF)$_2$SCN confirms the charge ordering with this charge (spin) configuration.~\cite{Fujiyama2004}

In the $P$-$T$ phase diagram, (TMTTF)$_2$AsF$_6$ and (TMTTF)PF$_6$ are located at the low-$P$ end which indicates the narrowest bandwidth. The resistivities for both materials have minima at about 200 K, and the ground states are known to be spin-Peierls state.~\cite{Delhaes1979,Laversanne1984}
The charge disproportionation in the TMTTF CTS was clearly demonstrated for the first time in these two materials by the spectral and $1/T_1$ studies of $^{13}$C NMR in the paramagnetic state.~\cite{Chow2000,Zamborszky2002} In Refs.\ \citen{Chow2000} and \citen{Zamborszky2002}, two distinct lines in the $^{13}$C NMR spectra appear to split with roughly equal intensities below about 100 K. Correspondingly, two $1/T_1$ appear due to the existence of unequal charge densities. It is argued that $1/T_1$ for the charge accepting and donating sites show different temperature dependences, and moreover, $1/T_1$ for the charge rich sites becomes independent of temperature between 40 K and 80 K indicating one dimensional ``magnetic" correlations among spin 1/2's realized at the charge accepting sites.
It is also found that under hydrostatic pressure, the phase transition temperature to the charge ordered state ($T_{\textrm{CO}}$) steeply decreases against pressure, and $T_{\textrm{CO}}$ vanishes at pressure as low as 0.15 GPa. It is proposed that the low temperature electronic state at ambient pressure is a coexistence of the charge ordered and the spin-Peierls states, which is fragile under pressure.

In this paper, we demonstrate results of the frequency shifts and nuclear spin lattice relaxation rate $1/T_1$ for the $^{13}$C labeled (TMTTF)$_2$AsF$_6$ from room temperature down to 2 K. Below 102 K, the splitting of the NMR spectra and the appearance of two $1/T_1$ with different temperature dependences evidence the charge ordering, which are consistent with the previously reported data in Refs.\ \citen{Chow2000} and \citen{Zamborszky2002}. We have successfully measured the ratio of $1/T_1$ of the charge accepting and donating sites down to the lowest temperature, for the first time, and pursued the development of the charge ordering. This ratio, suggesting the ratio in the charge densities to be 2:1, finally saturates in approaching $T_{\textrm{SP}}$ in contrast to the previous data above 30 K,~\cite{Zamborszky2002} which reveals only one spin correlation function even in the charge ordered state. However, at $T_{\textrm{SP}}$ the line splitting disappears, which we argue to originate from the variation in \textit{charge} densities. This shows a strong suppression of charge ordering or a redistribution of charge densities due to the phase transition from the charge ordered paramagnetic state to the spin-Peierls state. This reveals a qualitative difference from the proposed phase diagram based on high-field NMR measurements where the low temperature electronic state is translated from spin-Peierls state to the incommensurate SDW state. We argue possible origin to resolve the discrepancy and propose the electron-lattice coupling as the key interaction to suppress the charge ordering.

\section{Experiments}
\label{sec:Exp}
A rectangular needle-like single crystal of (TMTTF)$_{2}$AsF$_6$ in which the two central carbon sites on the TMTTF molecules are labeled with $^{13}$C was prepared by standard electrochemical oxidation method.

The uniform susceptibility ($\chi$) of our sample linearly decreases from 6$\times$10$^{-4}$ [emu/mol] at 300 K to 3.7$\times$10$^{-4}$ [emu/mol] at 20 K. However the $\chi$ shows a strong drop at 14 K and the electronic state undergoes a spin-Peierls state.

The NMR measurements were performed by using a standard pulsed NMR spectrometer operated at about 87.1 MHz with the bandwidth of 300 kHz for a single crystal. The spectra were obtained by the Fourier transformation (FT) of the spin echo refocused by applying $\pi/2$ and $\pi$ rf-pulses with the time interval of 20$\mu$s. The nuclear recovery data were obtained by integrating the intensities of distinct lines of the spectra in the frequency domain by saturation recovery method. The $1/T_1$ were defined by fitting them to a single exponential formula.

\section{Paramagnetic State}
\label{sec:Para}
\subsection{Angular dependence of the spectra for the present experimental setting}
\label{subsec:YokoRotate}
The $^{13}$C NMR spectra and nuclear relaxation shown in the following are obtained by applying rf-field along the needle direction that is nearly perpendicular to the TMTTF molecular planes. 

\begin{figure}[htb]
\centering
\includegraphics*[width=7cm]{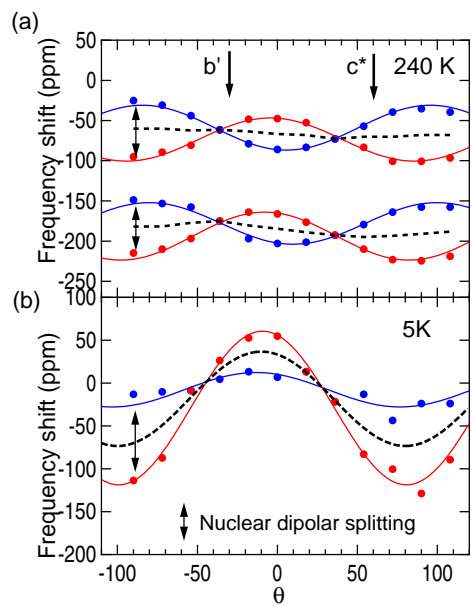}
\caption{(Color Online) Angular dependence of the peak positions of the NMR spectra rotated in the $b$-$c$ plane at 240 K (a) and  5 K (b).}
\label{fig:YokoRotateFig}
\end{figure}

We show here the angular dependence of peak frequencies of the spectra under rotation of magnetic fields in the $b$-$c$ plane. In this condition, the magnetic field is always applied along the molecular planes. There are two inequivalent $^{13}$C sites in the unit cell, and two lines are doubly split originating from nuclear dipolar interaction as shown in Fig.~\ref{fig:YokoRotateFig} (a). The sources of the angular dependence of the midpoints of the peak frequencies shown as dashed lines are the Knight (spin) shift ($K_s$) and chemical shift ($\sigma$). One can find small angular dependence of the midpoints of which amplitudes are less than 20 [ppm], although sinusoidal angular dependence $(3\cos ^2 \theta-1)$ are expected for both the spin and chemical shifts. This small angular dependence of $K_s+\sigma$ prevents from estimation of spin densities on carbon sites by plotting spin shifts against uniform susceptibility $(K$-$\chi$ plot).

The electronic state of (TMTTF)$_2$AsF$_6$ below 12 K is known to be the spin-Peierls state as mentioned above, so the spin shift is negligible in Fig.~\ref{fig:YokoRotateFig}~(b). Therefore, the sources for the angular dependence of the spectra at 5 K should be the anisotropy  in the chemical shifts and that in the nuclear dipolar interaction. In Fig.~\ref{fig:YokoRotateFig}~(b), the contribution from the nuclear dipolar interaction is indicated as vertical arrows, which have the same angular dependence as those in the paramagnetic state. The angular dependence of the chemical shift is therefore obtained by plotting the midpoints between the two peaks, as shown as the dashed line in Fig.~\ref{fig:YokoRotateFig}~(b).

This angular dependence can be ascribed to the in-plane anisotropy of the chemical shifts of the TMTTF molecules. The anisotropy of the chemical shift for TMTTF molecule can be speculated from that for BEDT-TTF molecule, since the local bonds connected to the central carbons in the TMTTF and BEDT-TTF molecules are similar to each other. 
For BEDT-TTF it is reported as (-24, 88, -64) [ppm], where $z$ axis is defined as the direction perpendicular to the plane of the BEDT-TTF molecule, $x (y)$ axis is defined as the long (short) axis parallel to the molecular plane~\cite{DeSoto1995}. Our observation of the angular dependence of the chemical shifts that shows in-plane anisotropy of 110 [ppm] as shown in Fig.~\ref{fig:YokoRotateFig}~(b) shows good agreement with the reported anisotropy for BEDT-TTF molecule.

The small angular dependence of $K_s+\sigma$ at 240 K indicates that the spin shifts and the chemical shifts have similar amplitudes of the in-plane anisotropy with opposite signs. As for the anisotropy of the spin shifts, it is estimated as (-225, -290, 515) [ppm] for (TMTTF)$_2$Br, that shows no phase transition above 14 K, from $K$-$\chi$ plot analysis~\cite{Barthel1993}. This estimation includes $-225-(-290)=65$ [ppm] in-plane anisotropy, that has similar amplitudes and opposite signs to that of the chemical shifts, which is consistent with above arguments.

In this section, we argued considerable reduction of the spin shift ($K_s$) by the chemical shift when applying external field normal to the molecular stacking axis, which prevents from the estimation of spin densities on central carbons on TMTTF molecules by the frequency shifts. An estimation of spin densities on molecules from the spin shifts by rotating external fields along $c$ axis is given in the Appendix.

\subsection{Line splitting of the spectra}
\label{subsec:ShiftPara}

\begin{figure}[htb]
\centering
\includegraphics*[width=7cm]{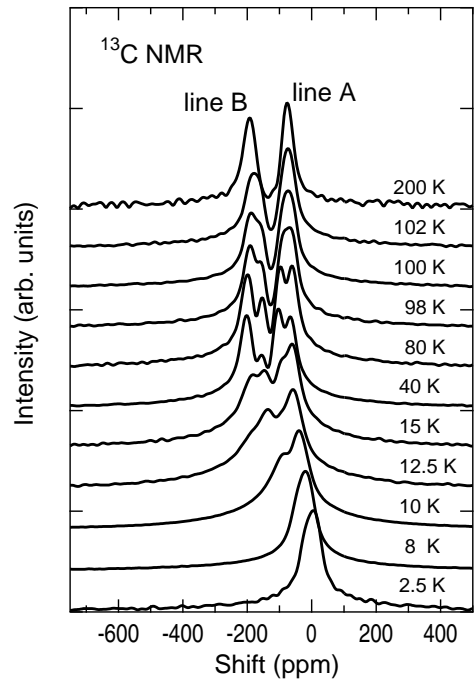}
\caption{The $^{13}$C NMR spectra at various temperatures. We label distinct lines as lines $A$ and $B$ from the peak at the highest frequency above 102 K, and as $A1$, $A2$, $B1$, and $B2$ between 15 K and 102 K.}
\label{fig:spectra}
\end{figure}

\begin{figure}[htb]
\centering
\includegraphics*[width=7cm]{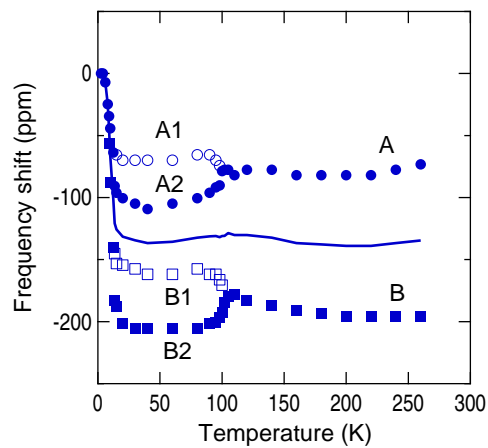}
\caption{The peak positions of the spectra. We define the peak position of the spectra at 2.5 K as the origin. The solid line is the averaged frequency of two (four) peak positions.}
\label{fig:Shift}
\end{figure}

The two $^{13}$C sites in the TMTTF molecule are crystallographically inequivalent in the crystal, which are the inner and the outer carbons from the inversion center. When external magnetic field is applied along the magic angle where doubly split lines due to the nuclear dipolar interaction overlaps, we obtain $^{13}$C NMR spectra as shown in Fig.~\ref{fig:spectra} composed of two distinct lines, $A$ and $B$, at 200 K. Each distinct line split into two peaks below 102 K, that is consistent with the reported data discussed to be the evidence for the charge ordering.~\cite{Chow2000} In Fig.~\ref{fig:Shift}, we plot the temperature dependences of the peak positions from the shift origin defined as the peak frequency at the lowest temperature. Each line above 102 K symmetrically splits at $T_{\textrm{CO}}$, so that the averaged frequency of the two (four) peaks of the spectra plotted as solid line is unchanged at this temperature. This shows that the total spin density is conserved at $T_{\textrm{CO}}$. The line splitting takes place between 102 K and 95 K. Below 95 K, four peaks are nearly independent of temperature.

\subsection{Nuclear spin lattice relaxation in the paramagnetic state}
\label{subsec:T1Para}
\begin{figure}[htb]
\centering
\includegraphics*[width=7cm]{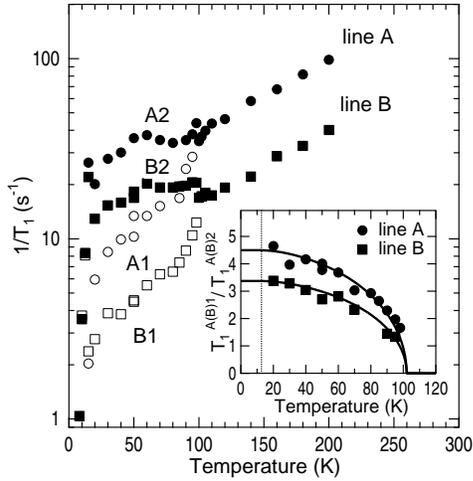}
\caption{$1/T_1$ for distinct lines. The ratios $T_1^{A1(B1)}/T_1^{A2(B2)}$ in the charge ordered state are plotted in the inset.}
\label{fig:T1}
\end{figure}

The nuclear spin lattice relaxation rate, $1/T_1$ for the distinct lines are shown in Fig.~\ref{fig:T1}. The $1/T_1$ for line $A$ ($1/T_1^A$) and that for line $B$ ($1/T_1^B$) both increase with temperature in a parallel manner above $T_{\textrm{CO}}$ in a logarithmic scale. The $1/T_1$ at site $n$ is expressed as, $1/T_1^n=\frac{2\gamma_n^2T}{\mu_B^2}\sum_q \, ^n F_\perp(q)^2 \chi''(q,\omega_L)/\omega_L$, where $\gamma_n$ is the gyromagnetic ratio of carbon, $^n F_\perp(q)$ is the $q$ dependent hyperfine coupling constant at site $n$, $\chi''$ is the imaginary part of the dynamic susceptibility, and $\omega_L$ is the Larmor frequency. Thus, the only difference between $1/T_1^A$ and $1/T_1^B$ is ascribed to that of $^n F_\perp(q)$. Here $F_\perp(q)$ is expected to have little $q$ dependence in organic conductors, therefore exclusively depends on the spin density at the nucleus. Since $1/T_1^A$ appears twice as large as $1/T_1^B$ above $T_{\textrm{CO}}$, we conclude that  line $A$ ($B$) in the frequency domain is due to the inner (outer) carbon site in the TMTTF molecule since the spin density on the inner carbon is estimated to be about 1.4 times as that on the outer one.~\cite{Kinoshita1984,Ishiguro1998}

Below $T_{\textrm{CO}}$, while $1/T_1^{A1(B1)}$ (the $1/T_1$ for the inner (outer) carbon in the charge donating molecules) decreases upon cooling, $1/T_1^{A2(B2)}$ (that for the charge accepting molecules) first increase and then turn to decrease. This fact evidences that the line splitting of the NMR spectra at 102 K at least originates from the difference of charge (spin) densities since the dominating source to the nuclear relaxation process here is expected to be the spin correlation function. It should be mentioned that the charge disproportionation takes place gradually from 102 K to about 60 K, although the line splitting occurs within a narrow temperature range of 5 K. This difference indicates that the contribution of the chemical shift to the frequency shift is larger than that of the spin shift at $T_\mathrm{CO}$.

The previously reported $1/T_1$ for the charge accepting sites ($1/T_1^{A2(B2)}$ in this paper) is nearly independent of temperature between 80 K and 40 K argued to be due to low-dimensional ``magnetic" correlations among spin 1/2's realized by the charge ordering.~\cite{Zamborszky2002} In the present study, however, $1/T_1^{A2(B2)}$ decreases from 60 K to 20 K in a nearly parallel manner in a logarithmic scale to that for the charge donating sites.

We plot the ratio $T_1^{A1(B1)}/T_1^{A2(B2)}$ in the inset of Fig.~\ref{fig:T1}. The ratio shows rapid growth just below $T_\mathrm{CO}$ but finally saturates in approaching $T_\mathrm{SP}$, showing the development of charge ordering. Therefore, we conclude that both the charge accepting and donating carbons follow one spin correlation function, $\chi''(q, \omega)$, and that the charge ordering discussed here seems to open a (pseudo) gap in $F(q)$ that is proportional to charge densities on the charge accepting and donating sites. Since both the charge accepting and donating sites follow the same correlation function, the ratio equals $(^{A(B)2}F_{\perp}(q)/^{A(B)1}F_{\perp}(q))^2$. Assuming that $F_{\perp}(q)$ is exclusively proportional to the charge density, it is derived that the ratio of the charge densities is about 2:1.

\section{Spin-Peierls State}
\label{sec:sP}
\begin{figure}[htb]
\centering
\includegraphics*[width=7cm]{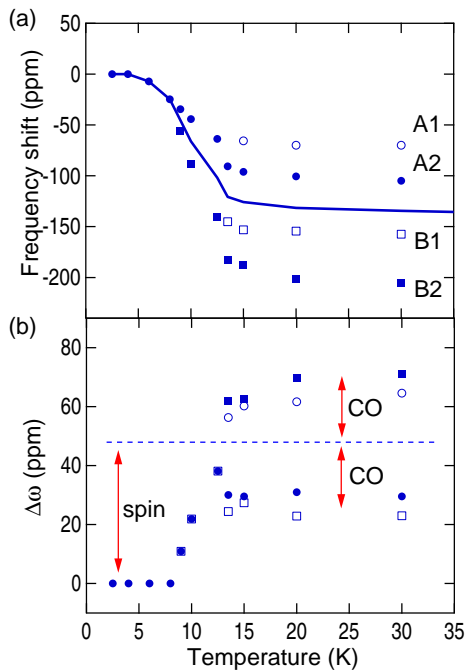}
\caption{(a) The peak positions in the vicinity and below $T_{\textrm{SP}}$. (b) The gaps between the peak positions and the averaged frequency denoted shown as the solid line in (a). The dashed line shows the spin contribution to $\Delta\omega$.}
\label{fig:ShiftLT}
\end{figure}

\begin{figure}[htb]
\centering
\includegraphics*[width=7cm]{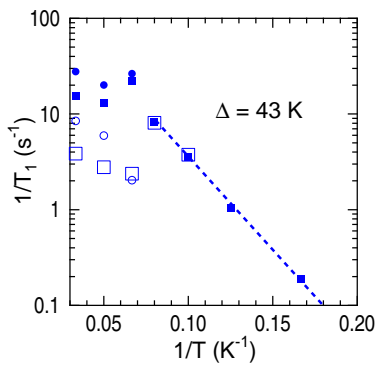}
\caption{$1/T_1$ for the distinct lines below 30 K. The dashed line shows the activated temperature dependence with $\Delta$ = 43 K.}
\label{fig:T1LT}
\end{figure}

Here we focus on the electronic state in the vicinity and below the spin-Peierls phase transition, as summarized in Figs.~\ref{fig:ShiftLT} and~\ref{fig:T1LT}.
As shown in Fig.~\ref{fig:ShiftLT}~(a), the strong cusp at 14 K for the averaged frequency (solid line) and the pronounced loss of spin density below this temperature shows the phase transition to the spin singlet ground state.

It is noteworthy that only two distinct lines are visible in the NMR spectra below $T_{\textrm{SP}}$. If the charge ordering have survived in the spin-Peierls state upon cooling, four lines would  be observed with monotonous decrease of the interval frequency between the two lines originated from charge accepting and donating carbons. 

In Fig.~\ref{fig:ShiftLT}~(b), we plot the difference between each peak positions and the averaged frequency denoted as the solid line in Fig.~\ref{fig:ShiftLT}~(a),  ($\Delta\omega \equiv |\nu_{\mathrm{res}, i}-\frac{1}{4}\Sigma^{4}_{i=1}\nu_{\mathrm{res}, i}|/\nu_0$). Here, $\nu_\mathrm{res}$ is the peak frequency of the spectra and $\nu_0$ is the Larmor frequency. Even in the paramagnetic state, while two $\Delta\omega$ for the charge rich sites slightly decrease two $\Delta\omega$ for the charge poor sites increase from 20 K to 14 K. The fact that $\Delta\omega$ at 12.5 K is about the averaged value of the four $\Delta\omega$ at 14 K excludes the possibility that the merging of the distinct lines is originated from loss of the total spin density across the spin-Peierls transition.

Likewise, $1/T_1$ for the charge donating and accepting sites merge into an intermediate value in between the separated values at 14 K as shown in Fig.~\ref{fig:T1LT}. This suggests that two inequivalently charged molecules below $T_{\textrm{CO}}$ again merge at $T_{\textrm{SP}}$.
$1/T_1$ below $T_{\textrm{SP}}$ well follows an activated temperature dependence ($1/T_1 \propto \exp(-\Delta /T)$) as shown in Fig.~\ref{fig:T1LT}, and yields $\Delta$=43 K as the activation energy. 

In addition to the spin contribution to our observations, the chemical shift that is originated from the motion of electrons on the molecular orbitals is reported to be affected by charge ordering.~\cite{Miyagawa2000} The lowest temperature $^{13}$C NMR spectra for $\theta$-(BEDT-TTF)$_2$RbZn(SCN)$_4$ that has spin-singlet ground state have two distinct peaks with the splitting of 80 ppm. Since the chemical shift is expected to be unchanged through the spin-Peierls transition, the only source of the line splitting is the differentiation in the chemical shift caused by inequivalently charged molecules. Therefore, again, our observation of only one peak in the NMR spectra at the lowest temperature indicates the absence or the strong suppression of the charge ordering in this material.

\section{Discussion}
\label{sec:Discussion}
We have confirmed charge ordering by the doubly split NMR spectra and corresponding differentiation in $1/T_1$ as shown in Sec.~\ref{sec:Para} below 102 K. This is consistent with the previously reported results of $^{13}$C NMR on this material above 30 K.~\cite{Zamborszky2002} Reference~\citen{Zamborszky2002} argues a possible scenario that one-dimensional ``magnetic'' correlation among spin 1/2 realized at the charge accepting molecules leads to the spin-Peierls transition, based on their results of $1/T_1$ that becomes independent of temperature between 40 K and 80 K. This argument naturally leads to a coexistence of charge ordered and spin-Peierls states that is theoretically reproduced based on a one-dimensional Hubbard model with lattice distortions.~\cite{Kuwabara2003} They also proposed a $P$-$T$ phase diagram including a coexistence of charge ordered and spin-Peierls state below 17 K and 0.15 GPa. 
This coexistence is indicated by the NMR spectra with ledges for (TMTTF)$_{2}$PF$_{6}$ under high magnetic fields where low temperature spin-Peierls state is suppressed and incommensurate SDW state is stabilized. 

In contrast to the arguments in Ref.~\citen{Zamborszky2002}, the merging of the split spectra shown in Sec.\ \ref{subsec:ShiftPara} reveals a disappearance or a strong suppression of the charge ordering below $T_{\textrm{SP}}$. Since the averaged frequency of two peaks in the spectra shows considerable shifts above 5 K originated from thermally excited magnons across the spin gap even in the spin-Peierls state, the charge ordering should have been observed also in the present experiments. These observations seems to exclude the simple scenario that the development of spin singlet correlation among spin 1/2's that are located at the molecular positions provokes the spin-Peierls transition. In other words, the spin-Peierls instability predominantly induces the spin correlation between the charge accepting and donating molecules and reduces charge ordering. In addition, the fact that both $1/T_1$ for the charge accepting and donating molecules decrease in a parallel manner in approaching $T_\mathrm{SP}$, indicating only one spin correlation function, is consistent with above arguments.

Theoretically, it is widely accepted that the Coulomb repulsion among the electrons stabilizes the charge ordered state in a quarter-filled Q1D electronic system. 
In particular, the interplay between charge ordering and lattice instability can realize various patterns of charge ordering.~\cite{Riera2000,Ung1994,Clay2003,Seo1997,Kobayashi1997,Mazumdar2000,Mazumdar1999,Hirsch1983,Penc1994} The stable charge configurations are CDW or BOW (bond-order wave) with 4$k_F$ (or 2$k_F$) periodicity including their coexistences. Of these, it is pointed out that the charge modulations of which antinodes are located in between two molecular sites such as BOW or BCDW (CDW with bond ordering) do not require significant charge disproportionation on the molecules.~\cite{Clay2003,Ung1994,Mazumdar1999,Riera2000} Our observations of the suppression of charge ordering with 4$k_F$ periodicity across the spin-Peierls transition may be caused by considerable electron-lattice coupling. However, at the present stage, we can not conclude whether the charge ordering is suppressed or whether other charge configuration than 4$k_F$-CDW becomes stabilized in the spin-Peierls state, because charge ordering is not a necessary condition for spin-Peierls transition as indicated by DCNQI based organic conductor which is other example of Q1D quarter-filled system. In this system, although (DI-DCNQI)$_2$Ag which goes into antiferromagnetic state shows charge ordering transition well above $T_N$, (DMe-DCNQI)$_2$Ag which undergoes spin-Peierls transition shows no charge ordering.~\cite{Hiraki1998} In this material, the antiferromagnetic correlation which causes the spin-Peierls state with $2k_F$ periodicity is considered to be that among the spin 1/2 realized by dimerized molecules without charge ordering (``dimer Mott insulator'').~\cite{Yonemitsu1997,Yonemitsu2001} As the charge configuration or the spin correlation, this ``dimer Mott'' type mechanism toward the spin-Peierls state can not be excluded.

As the origin of the redistribution of charge densities at $T_{\textrm{SP}}$, we can cite two possibilities which focus on electronic origin only and the coupling between the electronic correlation and lattice degree of freedom. One is a competitive interplay between two patterns for the charge ordering which would suppress the charge ordering. While the 4$k_F$ periodicity which is stabilized by Coulomb repulsions, the spin-Peierls state has 2$k_F$ modulation. The tetramization of the molecules can vary the charge configuration, and particularly induce charge transfers between two dimerized molecules. The other argument focuses on an interplay between the electronic correlation and lattice instabilities. The variation in the lattice parameters due to the spin-Peierls transition vary the charge transfers or Coulomb repulsion, that can reduce on-site type (CDW type) charge ordering. 

The repulsive interplay between charge ordering and spin-Peierls instability is also argued based on the topology of $P$-$T$ phase diagram in Ref.~\citen{Zamborszky2002}. The charge ordering in this material is so fragile that charge ordered paramagnetic state vanishes under a pressure as low as 0.15 GPa, indicating considerable competitive effect of lattice instability against the electronic correlations.  Our observations of the merging of the NMR spectra and nuclear relaxation also shows a competitive interplay between electronic instabilities toward two phase transitions. However, the fact that the suppression/redistribution of charge ordering seems to occur in a first order transition as indicated in Fig.~\ref{fig:ShiftLT} and \ref{fig:T1LT} suggests that rather the electron-lattice coupling than the spin-singlet correlation is the key interaction for the redistribution of charge configurations.

\section{Conclusion}
\label{sec:conclusion}
We demonstrate the frequency shifts and $1/T_1$ of central $^{13}$C sites for the quarter-filled Q1D material (TMTTF)$_2$AsF$_6$ from the room temperature to 2 K. The split of the distinct lines of the NMR spectra and the appearance of two $1/T_1$ which show different temperature dependences below 102 K evidence the charge ordering. The ratio of two $1/T_1$ for the charge accepting and donating sites which grows from $T_{\textrm{CO}}$ in lowering temperature and finally saturates in approaching $T_{\textrm{SP}}$ suggests the development of charge ordering. This shows that there exists only one spin correlation function even in the charge ordered state and the ratio of charge ordering is 2:1. 

In the vicinity or below the spin-Peierls transition, however, the split lines of the NMR spectra again merge by a charge origin. This shows a strong suppression of the charge ordering or a redistribution of ordered charge densities at the spin-Peierls transition.

\begin{acknowledgments}
One of the authors (S.\ F.) is grateful to K.\ Yonemitsu, H.\ Seo and K.\ Kanoda for fruitful discussions. This work is supported by a Grant-in-Aid for Scientific Research on Priority Area, ``Novel Functions of Molecular Conductors under Extreme Conditions" from the Ministry of Education, Culture, Sports, Science and Technology, and by a Grant-in-Aid for Scientific Research (No.\ 13640375) from Japan Society for the Promotion of Science.
\end{acknowledgments}

\appendix*
\section{Angular Dependence of the Spectra along the $a$-$b$ Plane}
As we discuss in Sec.~\ref{subsec:YokoRotate}, the frequency shift is reduced because the chemical shift and the spin shift have opposite signs when the external fields are rotated in the $b$-$c$ plane since the TMTTF molecules are stacked along the $a$ axis. Therefore, this experimental condition prevents a precise estimation of the hyperfine coupling tensor for $^{13}$C in TMTTF CTS.

In this Appendix, we demonstrate the orientation dependence of the frequency shifts where the field direction is rotated from $H\parallel a$ to $H\parallel b$. In approaching the $a$-axis (molecular stacking axis), the spin contribution to the frequency shift is expected to be larger because the external fields becomes closer to perpendicular to the molecular plane.

\begin{figure}[htb]
\centering
\includegraphics*[width=7cm]{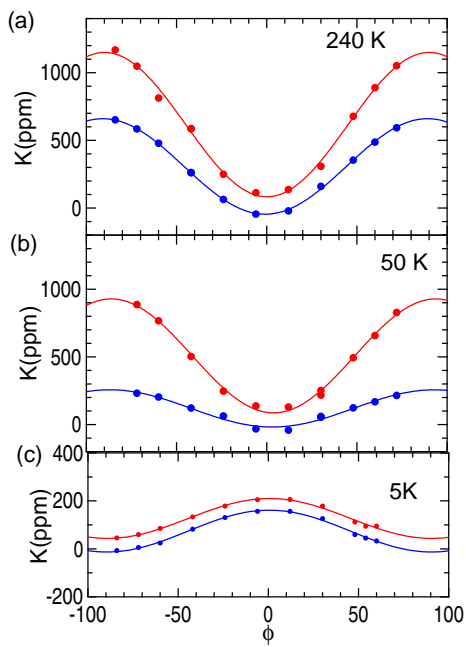}
\caption{Angular dependence of the frequency shifts along the $a$-$b$ plane at (a) 240 K, (b) 50 K, and (c) 5 K. At 240 (50) K, we plot only two peak positions although 4 (8) lines are observed. See the text.}
\label{fig:TateRotateFig}
\end{figure}

We plot the peak positions of the NMR spectra in Fig.~\ref{fig:TateRotateFig}. At 5 K, two lines are observed with a parallel gap along the rotation axis, which is due to the constant angle between the magnetic field and C-C doubled bond in the TMTTF molecule. The only source for the angular dependence is that of the chemical shift. At 240 K, we plot only two peaks out of four peaks that are located at the highest and lowest frequencies. Remaining two peaks stems from the nuclear dipolar interaction, which show parallel curves with two solid lines. At 50 K in the charge ordered state, we also plot two peaks in the highly complicated spectra. Here, 8 peaks are visible at most, that is consistent with the charge ordering as shown in Sec.\ \ref{subsec:ShiftPara}.

We summarize in Table \ref{tab:Fit} the fitted parameters of the peak positions in Fig.~\ref{fig:TateRotateFig} to the uniaxial function $\nu_{\mathrm{obs}}=\sigma_{\mathrm{iso}}+\sigma_{\mathrm{ax}}(1-3\cos^2\phi)+K_{\mathrm{iso}}+K_{\mathrm{ax}}(1-3\cos^2\phi)$. Here $\nu_{\mathrm{obs}}$ is the observed frequency, $\sigma_{\mathrm{iso}}(\sigma_{\mathrm{ax}})$ is the isotropic (anisotropic) part of the chemical shift, and $K_{\mathrm{iso}}(K_{\mathrm{ax}})$ is the isotropic (anisotropic) part of the Knight shift, respectively. We defined the shift origin as the isotropic part of the chemical shift. 

\begin{table}[htb]
\caption{\label{tab:Fit}Fitted parameters for the angular dependence of the peak frequencies of the NMR spectra for Fig.~ \ref{fig:TateRotateFig}. The shift origin is defined by the isotropic component of the chemical shift. The unit is ppm.}
\begin{tabular}{crr}
 &isotropic&anisotropic\\
\hline
$\sigma$& 0 & -56.7 \\
$K_{\textrm{s}}^1$& 68.8 &302.5\\
$K_{\textrm{s}}^2$& 284.2 & 410.2\\
\end{tabular}
\end{table}

The obtained spin contributions to the frequency shifts are more than 3 times as large as those in the $b$-$c$ plane, therefore we can estimate the ratio of the charge densities in the charge ordered state by the splitting of the NMR spectra. Since we can consider that the ratio of the spin densities for outer and inner carbons in the same molecule is unchanged across the charge ordering, the spectra at 50 K as shown in Fig.~\ref{fig:TateRotateFig} are located between the distinct lines stemmed from the outer carbon in the charge donating molecule and inner carbon in the charge accepting molecule. By comparing the angular dependence of the spectra at 50 K and 240 K, we obtain the ratio as  $(K^1_{\mathrm{ax}}/K^2_{\mathrm{ax}}(50 \mathrm{K}))/(K^1_{\mathrm{ax}}/K^2_{\mathrm{ax}}(240 \mathrm{K}))=2.2$. This value is consistent with the estimation by $1/T_1$ in Sec.\ \ref{subsec:T1Para}.


\end{document}